\documentclass[twocolumn,twoside,slac_two]{revtex4}
\usepackage{graphicx}
\usepackage{fancyhdr}
\begin{document}

\title{\huge Hydrodynamic Turbulence in Accretion Disks   }


\author{ Banibrata Mukhopadhyay, Niayesh Afshordi, Ramesh Narayan}
\affiliation{Harvard-Smithsonian Center for Astrophysics, 60 
Garden Street, MA 02138, USA}
%
%\author{P. Lucas}
%\affiliation{FNAL, Batavia, IL 60510, USA}

\begin{abstract}

Turbulent viscosity in cold accretion disks is likely to be
hydrodynamic in origin. We investigate the growth of hydrodynamic
perturbations in a small region of a disk, which we model as a linear
shear flow with Coriolis force, between two parallel walls. Although
there are no exponentially growing eigenmodes in this system, because
of the non-normal nature of the modes, it is possible to have a large
transient growth in the energy of certain perturbations. For a
constant angular momentum disk, the energy grows by more than a factor
of $1000$ for a Reynolds number of only $1000$, and so turbulence is
easily excited. For a Keplerian disk, the growth is more modest, and
energy growth by a factor of $1000$ requires a Reynolds number of
nearly a million. Accretion disks have even larger Reynolds numbers
than this. Therefore, transient growth of perturbations could seed
turbulence in such disks.

\end{abstract}

\maketitle

\thispagestyle{fancy}


\section{INTRODUCTION}

The origin of hydrodynamic turbulence is still not completely
understood.  Many efforts have been devoted to solve this problem,
beginning with the work of Kelvin, Rayleigh, and Reynolds at the end
of the nineteenth century, to more recent work.  In most cases, a
significant mismatch has been found between the results based on
linear theory and experimental (observational) data.  In the
laboratory, plane Poiseuille and plane Couette flows become turbulent
at Reynolds numbers, $R\ge R^{\rm exp}_c\sim 1000$, $350$, respectively,
while theoretically, plane Couette flow is linearly stable at all
values of $R$ and plane Poiseuille is linearly unstable only for
$R\ge R^{\rm the}_c=5772$.  What is the reason for this mismatch?  It would
appear that for $R$ values in between $R=R^{exp}_c$ and $R=R^{the}_c$,
some new effect allows the system to make a {\it subcritical}
transition to turbulence in the absence of a linear instability.

In the field of astrophysics, hydrodynamics and turbulence find
extensive applications. Accretion flows in binary stars, young stellar
objects and active galactic nuclei must be turbulent in order for the
gas to accrete. However the actual origin of the turbulence is unclear
and still under debate.  Inward mass accretion in a disk occurs as a
result of the transfer of angular momentum outward by means of some
viscous torque in the system.  More than three decades ago, Shakura \&
Sunyaev \cite{ss} and Lynden-Bell \& Pringle \cite{lp} proposed that
{\it turbulent} viscosity provides the necessary viscous torque to
cause inward mass transport.  However, the physical origin of the
turbulence in an accretion disk was unclear until the work of Balbus
\& Hawley \cite{bh} who showed that magnetized disks have a
Magneto-Rotational-Instability (MRI) \cite{veli,chan} in the
presence of a weak magnetic field.  This instability provides a
natural means to generate turbulence and transport angular momentum
outward. Later, Hawley, Gammie \& Balbus \cite{hgb1} showed that the
turbulence dies out if the Lorentz forces are turned off in a
magnetohydrodynamically turbulent Keplerian disk. Subsequently they
again showed \cite{hgb2} that the magnetic field dies out when the
tidal and Coriolis forces are turned off, keeping the Lorentz forces.
In two subsequent papers \cite{bhs,hbw}, it was shown through
numerical simulations that, whereas pure hydrodynamic turbulence is
easily triggered in plane Couette flow and in a constant angular
momentum disk, turbulence does not develop in an unmagnetized
Keplerian disk even in the presence of large initial perturbations.
The authors argued on this basis that hydrodynamic turbulence cannot
contribute to viscosity in accretion disks.

Despite the above results, it is probably premature to rule out
hydrodynamic turbulence in rotationally supported disks.  Many
astrophysical systems are known, e.g., star-forming disks, cataclysmic
variables in quiescence, outer regions of AGN disks, etc., in which
the gas is cold and neutral so that there is negligible coupling
between the magnetic field and the gas (see e.g., \cite{bb,gm,ftb}).
In these systems, the Magneto-Rotational-Instability (MRI) cannot play
a role.  How do these systems sustain mass transfer in the absence of the MRI?
What drives their turbulence?  It is possible that the turbulence is
purely {\it hydrodynamic} in origin.  Even in the absence of
exponentially growing modes, plane Couette flow is known to exhibit a
large transient growth in energy for certain initial conditions. Is
this transient growth responsible for the subcritical turbulence that
occurs in this flow, and could a similar phenomenon be responsible for
generating hydrodynamic turbulence in cold accretion disks?
Recent laboratory experiments on rotating Couette flow in the narrow
gap limit with linearly stable rotational angular velocity profiles
(similar to Keplerian disks) seem to indicate that turbulence does manage to
develop in such flows \cite{rz}. Longaretti \cite{long} points
out that the absence of turbulence in the simulations \cite{bhs,hbw}
may be because of their small effective Reynolds number.

Transient growth occurs in a linear system because of the non-normal
nature of the associated operator \cite{bf,rh,tt}.  How, and if, such
growth might finally cause turbulence is not well understood.  One
possibility is that the growth causes fluctuations to acquire large
amplitudes, and non-linear interactions then push the system into
self-sustained chaotic turbulence.

We present here some results on shearing rotating flows 
following the methods that have been developed to study the evolution
of non-normal modes in plane Couette flow.  We employ a local
approximation, and focus mainly on an eigenvalue analysis of
perturbations in Eulerian coordinates (for details, see \cite{man}, and
also \cite{amn,bgam} for a similar analysis in Lagrangian coordinates).  In
the next section, we briefly describe the basic model and write down
the set of equations to be solved. Then in \S 3, we present a few
important results.  We conclude with a discussion in \S 4.

\section{MODEL}

\begin{figure*}[t]
\centering
\includegraphics[width=50mm,angle=270]{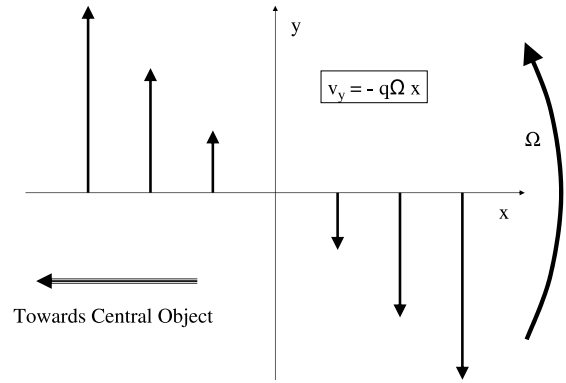}
\caption{Unperturbed flow in the local shearing box. The arrows represent
the velocity field.} \label{fig1}
\end{figure*}

We consider a small patch of an accretion disk whose local geometry is
described in terms of Cartesian coordinates, $x=(r-r_0)$,
$y=r_0(\phi-\phi_0)$.  We assume that the flow is incompressible and
extends from $x=-1$ to $+1$, between two rigid walls with {\it
no-slip} boundary conditions. The $y$ and $z$ directions are unbounded
with periodic boundary conditions.  This flow behaves like rotating
Couette flow in the narrow gap limit, when the unperturbed velocity
corresponds to a linear shear, $\vec{U}=(0,-x,0)$, and the Coriolis
force is described by the angular frequency $\vec{\omega}=(0,0,1/q)$:
$\Omega=\Omega_0(r_0/r)^q$.  Here all the coordinates and quantities
are expressed in terms of dimensionless variables.  Note, $q=3/2$
corresponds to a Keplerian disk, $q=1$ and $2$ to a flat rotation
curve and a constant angular momentum disk, respectively, and $q=0$
corresponds to rigid body rotation.  A schematic diagram of the flow
in the local coordinates is shown in Fig. \ref{fig1}.

Writing the perturbations as $U_x(x,y,z,t)\rightarrow u(x,y,z,t)$,
$U_y(x,y,z,t)\rightarrow U_y(x)+v(x,y,z,t)$, $U_z(x,y,z,t)\rightarrow
w(x,y,z,t)$, $P/\rho\rightarrow P/\rho+p(x,y,z,t)$, and combining the
linearized Navier-Stokes and continuity equations, we obtain
\begin{eqnarray}
%\nonumber
&&\hskip-1.cm\left(\frac{\partial}{\partial t}+U\frac{\partial}{\partial y}\right)\nabla^2 u
-\frac{\partial^2U}{\partial x^2} \frac{\partial u}{\partial y}
+\frac{2}{q}\frac{\partial \zeta}{\partial z}
=\frac{1}{R}\nabla^4 u,
\label{os}\\
&&\hskip-1.cm\left(\frac{\partial}{\partial t}+U\frac{\partial}{\partial y}\right)\zeta
-\frac{\partial U}{\partial x} \frac{\partial u}{\partial z}
-\frac{2}{q}\frac{\partial u}{\partial z}=\frac{1}{R}\nabla^2 \zeta,
\label{sq}
\end{eqnarray}
where $\zeta={\partial w}/{\partial y}-{\partial v}/{\partial z}$.
Equations (\ref{os}) and (\ref{sq}) are the standard Orr-Sommerfeld
and Squire equations, respectively, except that they now have
additional terms due to the presence of rotation in the system.  We
have to solve equations (\ref{os}) and (\ref{sq}) with the boundary
conditions: $u=\partial u/\partial x=\zeta=0$, at $x=\pm 1$ (no slip).

Due to translational invariance in the $y$ and
$z$ directions, the perturbations can be chosen as
$u(x,y,z,t)=\hat{u}(x)e^{i\sigma t}
e^{i(k_yy+k_zz)}$ and $\zeta(x,y,z,t)=\hat{\zeta}(x)e^{i\sigma t}
e^{i(k_yy+k_zz)}$, where $\sigma=\sigma_R+i\sigma_I$ is an eigenvalue
of the problem. After substituting these perturbations in (\ref{os})
and (\ref{sq}), we solve the combined Orr-Sommerfeld and Squire
equations to obtain the eigensystem
$\{\sigma,\hat{u},\hat{\zeta}\}$. The detailed set of equations and
its reduction to an eigenvalue equation form and the solution
procedure are described in \cite{man}.

As mentioned above, even in the absence of any exponentially growing
mode in the system, plane Couette flow can exhibit a large transient
growth in the energy of certain perturbations \cite{bf,tt}. This
growth occurs in the absence of non-linear effects and is believed to
facilitate the transition from laminar to turbulent flow.  The maximum
growth in the perturbed energy can be expressed as
\begin{eqnarray}
G_K(t)={\rm optimum}\left(\frac{E_t}{E_0}\right),
\label{grow3}
\end{eqnarray}
where
\begin{eqnarray}
E_t=\frac{1}{8k^2}\int_{-1}^{1}\left[k^2\hat{u}^\dag \hat{u}+\frac{\partial
\hat{u}}{\partial x}^\dag\frac{\partial \hat{u}}{\partial x}+\hat{\zeta}^\dag\hat{\zeta}\right]dx
\label{grow4}
\end{eqnarray}
and $E_0$ is the initial energy of the perturbation at time $t=0$.  By
``optimum" we mean that we consider all possible initial perturbations
and choose that function that maximizes the growth of energy at time
$t$. To understand the technical details of how to compute the growth,
see \cite{man}.

\section{RESULTS}

\begin{figure*}[t]
\centering
\includegraphics[width=100mm]{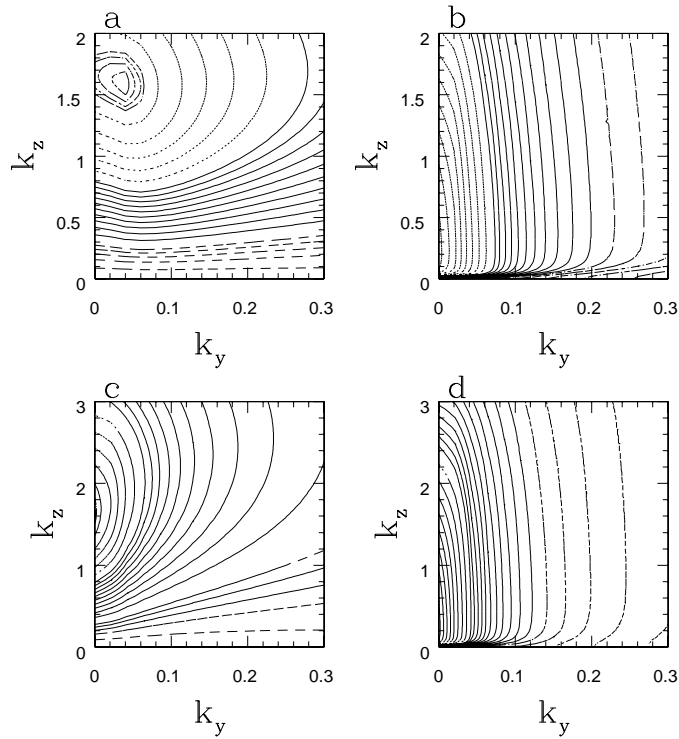}
\caption{Contours of (a) $G_{\rm max}$ for plane Couette flow, where
dashed contours correspond to $10,30,...90$; solid to $150,200,...600$;
dotted to $700,800,...1100$; dot-dashed to $1150,1160,...1200$;
(b) $t_{\rm max}$ for plane Couette flow, where dot-dashed contours correspond to  
$20,25,...45$; solid to $50,55,...100$; dotted to $110,120,...180$;
(c) $G_{\rm max}$ for $q=2$ flow, where contour labels are the same as in (a);
(d) $t_{\rm max}$ for $q=2$ flow, where contour labels are the same as in (b). 
All calculations are for $R=1000$.
} \label{fig2}
\end{figure*}

The eigenvalue analysis shows that, for plane Couette flow
and for a rotating flow with $q \le 2$, there are no
exponentially growing modes in the system, i.e., for no choice of the
parameters is there an eigenvalue with $\sigma_{I}>0$.  However, for
plane Couette flow, previous workers (e.g. \cite{bf,rh}) have shown
that there is a large transient growth even for $R=1000$.  The
interesting point is that the maximum growth in energy for
plane Couette flow and a constant angular momentum disk ($q=2$) are
very similar. Figure \ref{fig2} shows the contours of constant maximum
growth ($G_{\rm max}$) and the time at which the growth is maximized
($t_{\rm max}$) for both plane Couette and $q=2$. Whereas for $q=2$,
$G_{\rm max}$ occurs exactly on the $k_z$ axis, for plane Couette flow it
is slightly off the axis.  This is the only difference between the two
cases.

%\begin{table}[t]
%\begin{center}
%\caption{Margin Specifications}
%\begin{tabular}{|l|c|c|c|}
%\hline \textbf{Margin} & \textbf{Dual} & \textbf{A4 Paper} &
%\textbf{US Letter Paper}
%\\
%\hline Top & 7.6 mm & 37 mm & 19 mm \\
% & (0.3 in) & (1.45 in) & (0.75 in) \\
%\hline Bottom & 20 mm & 19 mm & 19 mm \\
% & (0.79 in) & (0.75 in)& (0.75 in) \\
%\hline Left & 20 mm & 20 mm & 20 mm \\
% & (0.79 in) & (0.79 in) & (0.79 in) \\
%\hline Right & 20 mm & 20 mm & 26 mm \\
% & (0.79 in) & (0.79 in) & (1.0 in) \\
%\hline
%\end{tabular}
%\label{l2ea4-t1}
%\end{center}
%\end{table}

\newpage
%\clearpage
%\begin{table*}[htbp]
\vskip0.2cm
{\centerline{\large Table 1}}
{\centerline{Maximum  Growth for Various $q$ and $R=1000$}}
\begin{center}
{
\vbox{
%\hspace {-2.2cm}
%\textwidth = 1.0in
%\psfig{width=2truecm}
%\textheight =8.0in
%\hskip -0.1cm
\begin{tabular}{ccccclllllllllllllll}
\hline
\hline
$q$ & $k_y$ & $k_z$ &  $G_{\rm max}$ & $t_{\rm max}$    \\
\hline
\hline
$1.5$  & $1.21$ & $0$ & $13.04$   & $8.8$  \\
\hline
$1.7$ & $1.06$ &$0.74$ & $13.4$   & $9$  \\
\hline
 $1.9$  & $0.54$ &$1.44$ & $23.75$   & $12.4$  \\
\hline
$1.99$  & $0.04$ &$1.8$ & $122.8$   & $27.3$  \\
\hline
 $2$  & $0$ &$1.66$ & $1165.9$   & $138.5$  \\
\hline
\hline
\end{tabular}
}}
\end{center}

\begin{figure*}[t]
\centering
\includegraphics[width=70mm]{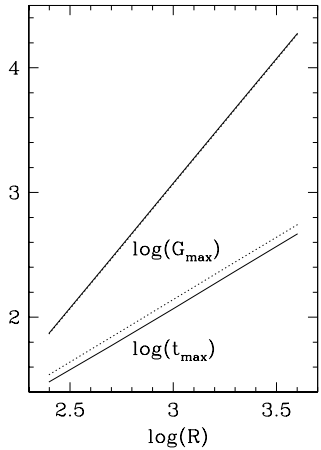}
\caption{Solid and dotted lines correspond to plane Couette and a $q=2$ disk respectively.
} \label{fig3}
\end{figure*}

Figure \ref{fig3} shows how $G_{\rm max}$ and $t_{\rm max}$ vary as a function
of Reynolds number, $R$, for plane Couette flow and $q=2$ flow. We see that
$G_{\rm max}$ scales as $R^2$ and $t_{\rm max}$ as $R$ in both cases, again
highlighting the similarity of the two flows.  The physical reason is
that a constant angular momentum disk has zero epicyclic frequency.
Thus, the effect of rotation is essentially cancelled out, making the
basic structure of the system very similar to that of plane Couette
flow.

\begin{figure*}[t]
\centering
\includegraphics[width=80mm]{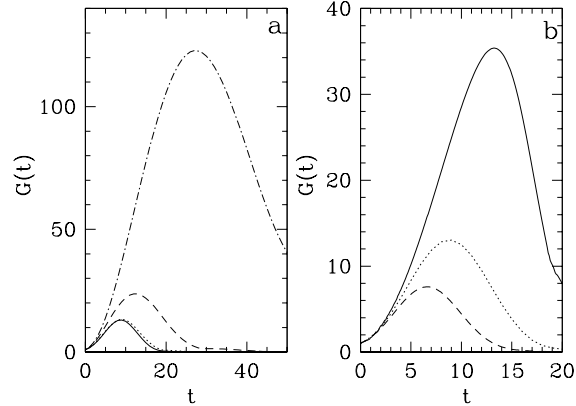}
\caption{Growth curves for (a) $q=1.5$ (solid), $1.7$ (dotted), $1.9$ (dashed),
$1.99$ (dot-dashed); other parameters are given in Table 1. (b) Growth curves for
$q=1.5$ when $\{R=4000,k_y=1.2\}$ 
(solid), $\{R=1000,k_y=1.21\}$ (dotted), $\{R=500,k_y=1.29\}$ (dashed); $k_z=0$.
} \label{fig4}
\end{figure*}

For a given $R$, when $q$ slightly deviates from $2$, growth
immediately reduces dramatically due to the introduction of a small
epicyclic frequency ($=\kappa=\sqrt{2(2-q)}\Omega$). This indicates
the dominant effect that $\kappa$ has on the fluid dynamics.  Table 1
and Fig. \ref{fig4}a show how $G_{\rm max}$ and $t_{\rm max}$ change as $q$
decreases below 2.  It is interesting to note that, for a $q=2$ flow
the growth is maximum for $k_y=0$, while for a Keplerian disk
($q=1.5$) the growth is maximum for $k_z=0$. Therefore, the location
of $G_{\rm max}$ moves in the $k_y-k_z$ plane systematically from one axis
to the other, as $q$ is varied.  Thus, for a constant angular momentum
disk, we need to include vertical structure in the perturbations to
understand the energy growth, whereas for a Keplerian disk a
2-dimensional analysis is sufficient.

Even though the growth decreases dramatically as $q$ falls below 2,
nevertheless, at a given $q$, if the Reynolds number $R$ increases to a
large enough value, the growth can become significant. In
Fig. \ref{fig4}b, we show how the peak value of growth increases with
increasing $R$ for Keplerian disks. It is found that $G_{\rm max}$ scales
as $R^{2/3}$ and $t_{\rm max}$ as $R^{1/3}$ for large $R$.  Therefore,
while the presence of a finite epicyclic frequency has a strong
dynamical effect on growth, it does not rule out a large growth. One
just needs much larger values of $R$.  Cold astrophysical accretion
disks can have $R$ as high as $10^{10}$ or more, so large growth
should in principle be easily achieved in such disks.  Also at smaller
$k_y$, $G_{\rm max}$ scales as $k_y^{2/3}$, while at larger $k_y$,
$G_{\rm max}$ decreases as $k_y^{-4/3}$. At large $t$, $t_{\rm max}$ scales as
$k_y^{-2/3}$. For a detailed derivation of these scaling relations, see
\cite{man}.

\begin{figure*}[t]
\centering
\includegraphics[width=70mm]{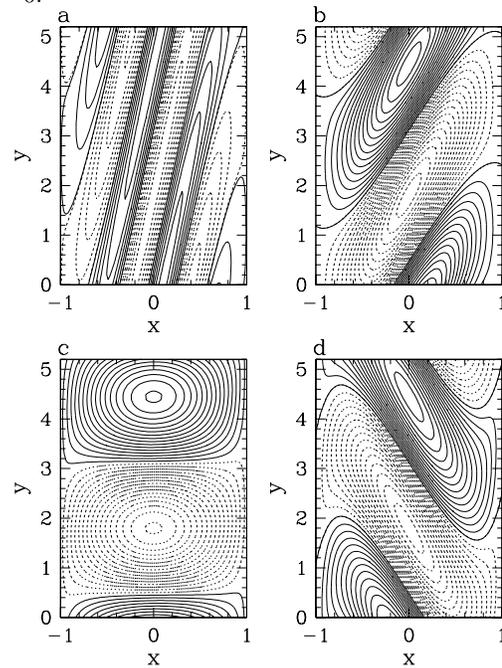}
\vskip1.1cm
\caption{Development of the perturbed $x$-component of the 
velocity $u(x,y)$ as a function of time 
for the optimal perturbation with the maximum growth of energy
in a Keplerian flow with $R=1000$. The 
perturbation has $k_y=1.21$, $k_z=0$, $t_{\rm max}=8.8$. Panels show the perturbation at
(a) $t=0$, (b) $t=t_{\rm max}/2=4.4$,
(c) $t=t_{\rm max}=8.8$, (d) $t=3t_{\rm max}/2=13.2$. Solid and dotted contours correspond to
positive and negative values of $u$ respectively.
} \label{fig5}
\end{figure*}

For a Keplerian flow the maximum growth occurs for $k_y\sim 1.2$,
$k_z=0$. Figure \ref{fig5} shows the development with time of the
perturbed velocity component $u(x,y)$ corresponding to $R=1000$. We
show snapshots corresponding to four times: $t=0$, $t_{\rm max}/2$,
$t_{\rm max}$, $3t_{\rm max}/2$.  The perturbations are seen to resemble plane
waves that are frozen in the shearing flow.  The initial perturbation
at $t=0$ is a leading wave with negative $x$-wavevector $k_x$ and with
$|k_x| \gg k_y$.  With time, the wavefronts are straightened out by
the shear, until at $t=t_{\rm max}$, the wavefronts are almost radial and
$k_x\sim0$.  Then, at yet later times, the wave becomes trailing and
the energy also decreases. The perturbations are very similar to the
growing perturbation described by other authors \cite{cg,tv,ur}.

\section{DISCUSSION}

We find that significant transient growth of perturbations is possible
in a shear flow with Coriolis force between walls (see also
\cite{yk}).  This system is an idealized local analog of an accretion
disk.  Although the system does not have any unstable eigenmodes,
nevertheless, because of the non-normal nature of the eigenmodes a
significant level of transient energy growth is possible for
appropriate choice of initial conditions. If the maximum growth
exceeds the threshold for inducing turbulence, it is plausible that
this mechanism could drive the system to a turbulent state.
Presumably, once the system becomes turbulent it can remain turbulent
as a result of nonlinear interactions and feedback among the
perturbations.

In this mechanism of turbulence, the maximum energy growth and the
time needed for this growth are probably the main factors that control
the transition to hydrodynamic turbulence.  As mentioned in \S 1, for
plane Couette flow $R_c^{exp}\sim 350$ and according to our analysis,
for $R=350$, $G_{\rm max}=145$, and the corresponding
$t_{\rm max}=42.3$. Since a $q=2$ disk is very similar to plane
Couette flow, the critical Reynolds number for turbulence in this
case is also likely to be $R_c\sim 350$.  For this $R$,
$G_{\rm max}=143.5$ and $t_{\rm max}=48.3$.

Based on the above idea, we make the plausible assumption that the
threshold energy growth factor needed for transition to turbulence in
a shear flow is $E_c \sim 145$.  Applying this prescription to the
optimal two-dimensional perturbations of a Keplerian disk (\S\S 2,3),
we estimate the critical Reynolds number for a Keplerian flow to be
$R_c\sim 3.4\times10^4$, i.e., a factor of 100 greater than in the case of
plane Couette flow.  The corresponding $t_{\rm max}=28.3$, which is
comparable to that in plane Couette flow, and is not too large
compared to the accretion time-scale of a geometrically thin disk.

Instead of taking $R_c\sim350$, we might wish to be conservative and
take $R_c\sim1000$ for plane Couette flow and $q=2$ flow.  In this
case, $G_{\rm max}\sim 1200$ and $t_{\rm max}\sim120-140$.  Applying the
requirement of $E_c\sim1200$ to a Keplerian flow, we find $R_c\sim 10^6$
and $t_{\rm max}\sim 100$.  Now the critical Reynolds number is a factor
of $1000$ greater than in the case of plane Couette flow.

An interesting result is that epicyclic motions in a
differentially-rotating disk kill the growth dramatically and as a
result the critical Reynolds number $R_c$ becomes higher. This also
changes the optimum wavevector $\{k_y,k_z\}$ of the perturbations
needed to produce energy growth.  For a constant angular momentum disk
($q=2$) and plane Couette flow, it is seen that growth is maximized
for $k_y\sim0$.  Even for a very small shift in the value of $q$ below
2, the location of maximum growth moves significantly in the $k_y-k_z$
plane away from the $k_z$ axis. With decreasing $q$, the epicyclic
motions of the disk increase, and correspondingly the optimum value
of $k_y$ for growth increases while the optimum $k_z$ decreases. For a
Keplerian disk, the growth is maximum for $k_z=0$ (on the $k_y$
axis). To the best of our knowledge, this change in the location of
the maximum growth in the $k_y-k_z$ plane has not been commented upon
prior to this work.

In earlier numerical simulations \cite{bhs,hbw}, the above reduction of growth due to
the effect of epicyclic motion in the disk was already
noticed. However, those authors then proceeded to rule out the
possibility of hydrodynamic turbulence in Keplerian disks.  We do not
agree with this conclusion.  As we have shown, Keplerian disks can
support large transient energy growth, only they need much larger
Reynolds numbers to achieve the same energy growth as plane Couette
flow or a $q=2$ flow.  The numerical simulations probably had
effective Reynolds numbers $\lesssim10^4$ (because of numerical
viscosity) which is below our most optimistic estimate of the critical
Reynolds number for a Keplerian flow.  Thus, we suspect the
simulations simply did not have sufficient numerical resolution to
permit the turbulence. The same point has been made by other authors
\cite{long,ur}.

We conclude with an important caveat.  While the demonstration of
large energy growth is an important step, it does not prove that
Keplerian disks will necessarily become hydrodynamically turbulent.
Umurhan \& Regev \cite{ur} have shown via two-dimensional simulations
that chaotic motions can persist for a time much longer than the time
scale $t_{\rm max}$ needed for linear growth.  However, they also note
that their perturbations must ultimately decline to zero in the
presence of viscosity.  To overcome this limitation, it is necessary
to invoke three-dimensional effects.  Secondary instabilities of
various kinds, such as the elliptical instability (e.g. \cite{ker}),
are widely discussed as a possible route to self-sustained turbulence
in linearly perturbed shear flows.  It remains to be seen if these
instabilities are present in perturbed flows such as those shown in
Figure \ref{fig5}.

% \begin{figure}
% \end{figure}

% Surround figure environment with turnpage environment for landscape
% figure
% \begin{figure}
% \end{figure}

% If in two-column mode, this environment will change to single-column
% format so that long equations can be displayed. Use
%\begin{widetext}
%\end{widetext}

\vskip1.0cm
\begin{acknowledgments}
This work was supported in part by NASA grant NAG5-10780 and NSF grant
AST 0307433.
\end{acknowledgments}

%\begin{thebibliography}{9}   % Use for  1-9  references

\end{document}